%% file: paper.tex
\documentclass[manuscript,anonymous=false]{acmart}
\usepackage{arydshln}
\usepackage{graphicx}
\usepackage{times}

\usepackage{soul}
\usepackage{url}
\usepackage{graphicx,subcaption}
\usepackage{amsmath}
\usepackage{booktabs}
\usepackage{bbm}
\usepackage{multirow}
\usepackage[ruled,vlined,linesnumbered]{algorithm2e}
\usepackage{etoolbox}
\usepackage{enumitem}
\makeatletter
\patchcmd\algocf@Vline{\vrule}{\vrule \kern-0.4pt}{}{}
\patchcmd\algocf@Vsline{\vrule}{\vrule \kern-0.4pt}{}{}
\makeatother
\newtheorem{problem}{Problem}
\AtBeginDocument{%
  \providecommand\BibTeX{{%
    \normalfont B\kern-0.5em{\scshape i\kern-0.25em b}\kern-0.8em\TeX}}}

\setcopyright{acmcopyright}
\copyrightyear{2018}
\acmYear{2018}
\acmDOI{10.1145/1122445.1122456}

\acmJournal{JACM}
\acmVolume{37}
\acmNumber{4}
\acmArticle{111}
\acmMonth{8}

\begin{document}

\title{Personalized TV Recommendation: Fusing User Behavior and Preferences}
\author{Sheng-Chieh Lin}
\affiliation{%
  \institution{Academia Sinica}
}
\authornote{Both authors contributed equally to this research.}
\email{jacklin_64@citi.sinica.edu.tw}
\orcid{1234-5678-9012}
\author{Ting-Wei Lin}
\authornotemark[1]
\email{jacky841114@gmail.com}
\affiliation{%
  \institution{National Chengchi University}
}

\author{Jing-Kai Lou}
\email{kaelou@kkstream.com}
\affiliation{%
 \institution{KKStream Limited}
}
\author{Ming-Feng Tsai}
\email{mftsai@nccu.edu.tw}
\affiliation{%
 \institution{National Chengchi University}
}

\author{Chuan-Ju Wang}
\email{cjwang@citi.sinica.edu.tw}
\affiliation{%
 \institution{Academia Sinica}
}

\renewcommand{\shortauthors}{Trovato and Tobin, et al.}

\begin{abstract}
In this paper, we propose a two-stage ranking approach for recommending linear
TV programs.
The proposed approach first leverages user viewing patterns regarding
time and TV channels to identify potential candidates for recommendation and
then further leverages user preferences to rank these candidates given textual
information about programs.
To evaluate the method, we conduct empirical studies on a real-world TV
dataset, the results of which demonstrate the superior performance of our model in
terms of both recommendation accuracy and time efficiency. 
\end{abstract}


\maketitle

\section{Introduction}
Linear TV programs play crucial roles in our daily lives.
With the quickly increasing number of TV channels and programs, it is
important to develop effective recommender systems for TV users.
Although the development of recommender systems has been stimulated by the
rapid growth of information on the Internet, and many algorithms have been
successfully applied to various online services (e.g., music and video
streaming services)~\cite{Gomez-Uribe16, Yang2018,Chen2019}, little has been done for personalized TV
recommendation (TV Rec) in the literature.
Most well-developed recommendation algorithms are not applicable for such a
recommendation problem due to the following two key challenges of TV Rec: (1)
Complete-item cold start: Unlike video on demand (VOD), new TV programs are
released on a daily basis (although some drama or movies are replayed, they
usually have different titles or descriptions);\footnote{Another practical
challenge is that the programs that share common content do not share an
identical ID, which rules out directly adopting collaborative filtering or matrix
factorization in real-world scenarios.}  
(2) Context awareness: user viewing behavior for TV programs strongly
depends on their conditions (e.g., time and mood); for instance, 
watching news during dinner but preferring sports in the morning.

To address the first challenge, some studies adopt content-based approaches combined with collaborative filtering (CF) for TV Rec~\cite{ali2004tivo,cotter2000ptv,fernandez2006avatar,Smyth1999,david2012}.
However, these approaches do not consider the second key characteristic---context awareness---in TV Rec, for which another line of work focuses mainly on characterizing users' time-aware preferences~\cite{ardissono2004user,turrin2014time,david2012,Yu17,Kim2018ATR}.
Although, these studies model users' time-aware preferences regarding channels and program genres, they do not precisely reflect users' viewing preferences regarding program content.
This is due to the fact that users' access to channels also depends on their viewing habits or location, and that genres are merely coarse-grained information about programs and thus provide little information about program content.
Moreover, ~\cite{Hsu2007} further accounts user moods but such user data is difficult to obtain and even harder to measure.

To address the above two challenges within a unified framework, we
propose a two-stage ranking approach for TV Rec which consists of two
components: one to model viewing behavior and the other for viewing
preferences.
Specifically, viewing behavior refers to users' viewing patterns regarding
time and TV channels, whereas viewing preferences refers to preferences
regarding the content of TV programs.
For the former, we adopt a finer granularity in terms of time than
previous work (e.g., days$\times$hours in \cite{ardissono2004user,turrin2014time}), whereas for the latter, we
leverage textual information about programs to better model user viewing
preferences.
Moreover, inspired by the capabilities and limitations of the two components, we
propose fusing them with a simple yet effective two-stage ranking algorithm
that locates potential candidates based on the first component and then
further ranks them based on the second component.
Also note that in the literature, this is the first work
to formally define the problem of TV Rec and provide a unified approach to
capture both user viewing behavior and preferences. 
Empirical results on a real-world TV dataset demonstrate its effectiveness in
recommendation; at the same time, this approach is advantageous and
practical for real-world applications due to its time-efficient and
parameter-free design.



\section{Methodology}
\label{sec:methodology}
\subsection{Problem Formulation}
Personalized TV recommendation (TV Rec) is the task of recommending
yet-to-be-released TV programs to a group of users. 
To properly formulate the problem and our proposed method, we first define
three terms: 1) weekly time slot, 2) interaction tensor, and 3) program meta
information required for TV Rec. With these definitions, we formalize
personalized TV Rec as a top-$k$ recommendation problem given user-implicit
feedback.  

\begin{definition}[{\bf Weekly time slot}]
A weekly time interval can be equally divided into $n$ weekly time slots, each
of which is denoted as $w_i=(t_i,t_{i+1}]$, where $t_i$ ($t_{i+1}$) denotes
the beginning time (the end time, respectively) of the $i$-th time slot.
Together, all of the time slots compose set $W=\{ w_i|1\leq i \leq n\}$.
Thus, any given timestamp $\mathbf{s}\in S$ can be projected onto a weekly time
slot $w_{\mathcal{T}(\mathbf{s})} \in W$ by function
$\mathcal{T}(\cdot):S\rightarrow\{1,\cdots,n\}$, where $S$ denotes a set of
arbitrary timestamps.
\end{definition}
For example, when we divide a week into 168 time
slots (i.e., one hour for each time slot), we have $W=\{w_1=[\text{Mon 00:00},
\text{Mon 01:00}),\cdots, w_{168}=[\text{Sun 23:00}, \text{Mon 00:00}) \}$,
in which the specific timestamp ``May 11, 2020, 05:30 (Mon)'' belongs to the
6th time slot, $w_6$.   
Note that a given time span $[\mathbf{s},\mathbf{e}]$ can also be projected onto
a set of time slots $\{ w_j |\mathcal{T}(\mathbf{s})\leq j\leq
\mathcal{T}(\mathbf{e}) \}$.
Also note that in our later empirical studies, we adopt a finer
granularity in terms of time (i.e.,~15 minutes as the length of the time slot) than prior art.

\begin{definition} [{\bf Interaction tensor}] Let $U$, $I$, and $C$ denote the sets
of users, TV programs, and TV channels, respectively. An interaction tensor,
denoted as $\mathcal{A}=(a_{u,i,w,c})\in \mathbb{R}^{|U|\times |I|\times
|W|\times |C|}$, represents user-item associations through a certain channel
within a certain weekly time slot, where $a_{u,i,w,c}$ denotes the weight of
the association. Note that the tensor is binary for implicit feedback;
that is, if user $u\in U$ views program $i\in I$ played in channel $c\in
C$ within time slot $w\in W$, $a_{u,i,w,c}=1$; otherwise, $a_{u,i,w,c}=0$.

\end{definition}

\begin{definition}[{\bf Program meta information}] 
\label{def:meta}
Given a set of TV programs $I$, meta information for each $i\in I$ records that
program $i$ is broadcast by channel $\mathrm{CH}({i}) \in C$ at the time
interval $[\mathbf{s}_i,\mathbf{e}_{i}]$ with the content information
$\mathrm{CNT}(i)$, where $\mathrm{CH}(\cdot)$ and $\mathrm{CNT}(\cdot)$ are the
projection functions respectively mapping program $i$ to its channel and its
textual information (e.g., title, artists, and abstract).
\end{definition}

\begin{problem}
\label{def:problem setting}
\textbf{Top-$k$ TV Recommendation from Implicit Feedback.} 
Let $I_{\rm train}$ and $I_{\rm test}$ denote the sets of TV programs broadcast
in the past (training data) and in the future (test data), respectively; note
that for the problem of TV Rec, $I_{\rm train}\bigcap I_{\rm
test}= \emptyset$.
Given a historical interaction tensor $\mathcal{A}_{\rm train}=(a_{u,i,w,c})
\in \mathbb{R}^{|U|\times |I_{\rm train}|\times|W| \times |C|}$, for each user
$u \in U$, we identify the top-$k$ programs from the set of yet-to-be-released 
(new) programs $I_{\rm test}$ by leveraging the information from
$\mathcal{A}_{\rm train}$ and meta information of $I_{\rm train}\bigcup I_{\rm
test}$. 
\end{problem}

\subsection{Proposed Method}
With a TV recommender system, we seek to leverage historical viewing logs and
content information of programs to infer two user characteristics: (1) behavior
and (2) preferences, which are addressed in Sections~\ref{sec:behavior}
and~\ref{sec:preference}, respectively. 
We then propose a simple yet effective two-stage ranking method in
Section~\ref{sec:algorithm} that takes into account both user characteristics,
thereby fusing user viewing habits and preferences into the modeling process.

\subsubsection{Viewing behavior}\label{sec:behavior}
Here, we define the so-called \emph{viewing behavior} of users
based on the following observations. 
As suggested by \cite{turrin2014time}, most TV users exhibit predictable viewing behavior
strongly connected to weekly time slots and TV channels.
Intuitively, users prefer to watch TV during their leisure time, which heavily
depends on their work and lifestyle. 
In addition, users tend to switch between a limited number of channels even
though they have a large number to choose from. 
Thus a user's TV viewing behavior can be defined as the
probability distribution of watching TV on a given channel at a given time.

Given a historical user-item interaction tensor $\mathcal{A}_{\rm
train}=(a_{u,i,w,c}) \in \mathbb{R}^{|U|\times |I_{\rm train}|\times|W| \times
|C|}$, we extract each user $\mathbf{u}$'s viewing behavior by computing his
or her viewing probability distribution over weekly time slots $W$ and TV
channels $C$. 
Formally speaking, we represent each $\mathbf{u}$'s viewing behavior as a
probability distribution matrix,
$\mathcal{B}^{\mathbf{u}}=(b^{\mathbf{u}}_{\mathbf{w},\mathbf{c}}) \in
\mathbb{R}^{\left| W \right| \times \left| C \right|}  $, where each element
$b^{\mathbf{u}}_{\mathbf{w},\mathbf{c}}$ is defined as
\begin{equation}
\label{eq:watching_behavior}
b^{\mathbf{u}}_{\mathbf{w},\mathbf{c}}=
\left(
\sum\limits_{i,w,c}a_{\mathbf{u},i,w,c} 
 \mathbbm{1}_{ \{ w=\mathbf{w} \} } 
 \mathbbm{1}_{ \{ c=\mathbf{c}\} } 
\right)\left/
\left(\sum\limits_{i, w, c }a_{\mathbf{u},i, w, c } \right)\right.
.
\end{equation}

Additionally, in order to recommend yet-to-be-released TV programs for users based
on their viewing behavior, we construct the matrix
$\mathcal{B}^{\mathbf{i}'}=(b^{\mathbf{i}'}_{\mathbf{w},\mathbf{c}}) \in
\mathbb{R}^{\left| W \right| \times \left| C \right|}  $ for each new item
$\mathbf{i}' \in I_{\rm }$ using the meta information defined in
Definition~\ref{def:meta}, where $b^{\mathbf{i}'}_{\mathbf{w},\mathbf{c}}=
\mathbbm{1}_{ \{ \mathbf{w} \in \{ w_j |\mathcal{T}(\mathbf{s}_{\mathbf{i}'})\leq j\leq \mathcal{T}(\mathbf{e}_{\mathbf{i}'}) \} \} } 
\cdot \mathbbm{1}_{ \{ \mathrm{CH}(\mathbf{i}')=\mathbf{c}\} } $.
Recall that $[\mathbf{s}_{\mathbf{i}'},\mathbf{e}_{\mathbf{i}'}]$ denotes the
time interval during which program $\mathbf{i}'$ is broadcast.
Finally, we compute the matching score between $\mathbf{u}$ and
$\mathbf{i}^{'}$ given viewing behavior as 
\begin{equation}
\label{eq:watching_behavior_score}
\begin{aligned}
s_{\mathbf{u},\mathbf{i'}}^{b}
=\mathrm{MAX}\left(\mathcal{B}^{\mathbf{u}} \odot \mathcal{B}^{\mathbf{i'}}\right) \text{ and } (w,c) = {\rm IdxMax}\left(\mathcal{B}^{\mathbf{u}} \odot \mathcal{B}^{\mathbf{i'}}\right),
\end{aligned}
\end{equation}
where $\odot$ denotes element-wise multiplication between two matrices,
$\mathrm{MAX}(\cdot)$ is the function to extract the maximum element in a
matrix, and ${\rm IdxMax}(\cdot)$ locates the indices of the maximum
element.\footnote{In practice, there is no need to conduct the element-wise
multiplication to get $s^{b}_{\mathbf{u},\mathbf{i}'}$; instead, for each
$\mathbf{i}'$, $s^{b}_{\mathbf{u},\mathbf{i}'}$ is the maximum in the set $\{b^{\mathbf{u}}_{\mathbf{w},\mathbf{c}}|
\mathbf{w} \in \{ w_j |\mathcal{T}(\mathbf{s}_{\mathbf{i}'})\leq j\leq
\mathcal{T}(\mathbf{e}_{\mathbf{i}'})\}\wedge \mathbf{c}={\rm CH}(\mathbf{i}') \}$.}
Note that $s_{\mathbf{u},\mathbf{i'}}^{b}$ is the estimated probability
that user $\mathbf{u}$ views item $\mathbf{i'}$ given his or her
historical viewing behavior.

\subsubsection{Viewing Preferences}\label{sec:preference}
In contrast to the aforementioned user behavior, a user's
preferences are usually associated with the content of his or her preferred
items. 
We formally define a user's \emph{viewing preferences} as the
program contents he or she prefers to watch, which we represent in the 
proposed method using the textual information of programs.
Note that as with a typical TV Rec scenario, all candidate items in $I_{\rm
test}$ for recommendation are new, which is the same as the complete cold-start
problem in typical recommender systems. 
Such a problem is commonly addressed using content-based approaches~\cite{david2012, Chou2016}; 
likewise, we here use textual item information
to locate new items for recommendation.

For each program $\mathbf{i} \in I_{\rm train}$, we map its content information
to a $d$-dimensional embedding~$h_{\mathbf{i}}$ using a text encoder~$\mathcal{E}$:
\begin{equation}
\label{eq:text_enc}
{h}_{\mathbf{i}}=\mathcal{E}\left(\mathrm{CNT}(\mathbf{i})\right) \in \mathbb{R}^{d}.
\end{equation}
In order to map user~$\mathbf{u}$'s preferences to the same embedding space, we
gather all the programs associated with $\mathbf{u}$ in the training data, after
which we compute the average pooling over their embeddings to obtain
$\mathbf{u}$'s viewing preferences~$h_{\mathbf{u}}$ as
\begin{equation}
h_{\mathbf{u}}= 
\frac{\sum_{i\in I^{\mathbf{u}}_{\rm train}} h_i}{|I_{\rm train}^{\mathbf{u}}|}  \in \mathbb{R}^{d},
\label{eq:gpreference}
\end{equation}
where $I^{\mathbf{u}}_{\rm train}= \{ i\,|\,i \in I_{\rm train} \wedge
\exists\, w\in W, c\in C\, a_{\mathbf{u},i,w,c}=1\}$.
Similarly, for each item $\mathbf{i}' \in I_{\rm test}$, we project its content
information using the same text encoder~$\mathcal{E}$ from Eq.~(\ref{eq:text_enc}). 
Finally, the matching score for $\mathbf{u}$ and $\mathbf{i}^{'}$ in terms of 
of viewing preferences is computed as
\begin{equation}
\label{eq:preference_score}
    s^{p}_{\mathbf{u},\mathbf{i}'}=\langle h_{\mathbf{u}}, {h_{\mathbf{i}'}}\rangle,
\end{equation}
where $\langle\cdot,\cdot\rangle$ denotes the dot product of two vectors.

In addition, for TV Rec, it is common that multiple users (i.e.,
family members) share the same account, under which these users may have
different viewing preferences and watch TV at different weekly time slots.
For example, whereas children enjoy watching cartoons after school, parents
prefer to watch news or dramas after work.
We address this by further tailoring the viewing preferences of an
``account'' to time-aware preferences; that is, for each 
account~$\mathbf{u}\in U$ and each time slot~$\mathbf{w}\in W$, we have
\begin{equation}
    h_{\mathbf{u}, \mathbf{w}}= 
\frac{\sum_{i \in I^{\mathbf{u}, \mathbf{w}}_{\rm train}}h_i}{|I_{\rm train}^{\mathbf{u}, \mathbf{w}}|}  \in \mathbb{R}^{d},
\label{eq:tpreference}
\end{equation}
where $I^{\mathbf{u},\mathbf{w}}_{\rm train}= \{ i\,|\,i \in I_{\rm
train}\wedge  \exists\,c\in C\, a_{\mathbf{u},i,\mathbf{w},c}=1\}$. 
With these fine-grained viewing preferences, the score of user~$\mathbf{u}$ for
item $\mathbf{i}'$ becomes 
\begin{equation}
\label{eq:time_preference_score}
    s^{p}_{\mathbf{u},\mathbf{i}'}=\langle h_{\mathbf{u},{w}_j}, {h_{\mathbf{i}'}}\rangle,
\end{equation}
where ${w}_{j}\in W$ denotes the time slot in which item $\mathbf{i}'$ begins
playing; i.e., $j=\mathcal{T}(\mathbf{s}_{\mathbf{i}'})$. 

\subsubsection{Two-stage Ranking}\label{sec:algorithm}

In this section, we propose a two-stage ranking approach that 
leverages the above two features---user viewing behavior and user viewing
preferences---for TV Rec. 
Before describing the proposed approach, we make observations and
lay out the motivation of our design based on the limitations of each feature as follows. 
\begin{itemize}[leftmargin=*]
    \item 
	 \textbf{Viewing behavior}: In practice, there are usually
	 multiple programs broadcast on the same channel at the same time slot; in
	 this case, these programs are given the same matching score for a user in terms 
	 of his or her viewing behavior. Thus, recommendation that is based solely
	 on user viewing behavior chooses all the programs from a
	 certain channel and time slot.\footnote{When multiple programs have the
	 same score, we assign a higher rank to programs with earlier starting times.}
	 However, in a real-world scenario, it is unlikely that a user at a
	 given time slot watches more than one TV program, especially for
	 short time slots;\footnote{In the experiments, we adopted 15 minutes as our
	 time slot interval, an optimal setting for using only viewing
	 behavior for recommendation; even in this case, each time slot nevertheless contains 1.5
	 programs on average.} in this case recommending multiple programs from the
	 same channel at a given time slot could lead to poor recommendation
	 quality.
    \item
	 \textbf{Viewing preferences}: Although user preferences are useful for
	 recommendation, recommending linear TV programs based solely thereon usually
	 results in low accuracy. For example, if an office worker enjoys watching
	 action movies during the weekend, it is unreasonable to recommend action
	 movies at midnight during weekdays.
\end{itemize}

Based on the above characteristics and limitations, we propose two-stage
ranking to leverage the two features for TV Rec, as
detailed in Algorithm~\ref{alg:ts_ranking}.
Briefly speaking, for each user~$\mathbf{u}$, we propose first ranking the
program set~$I_{\rm test}$ according to viewing 
behavior~($s^{b}_{\mathbf{u},\mathbf{i}'}$) (lines 2--6);  then, at the second stage
(lines 7--15), for those programs broadcast on the same channel at the same
weekly time slot, we choose only one program among them according to the user's
viewing preferences~($s^p_{\mathbf{u},\mathbf{i}'}$). 
Note that we put the model for viewing behavior at the first stage as previous studies indicate that the viewing behavior usually dominates the recommendation performance~\cite{turrin2014time}, which is also consistent with the finding in our later experiments.
This approach boasts two advantages: 1) it is
parameter-free, and 2) it is computationally efficient as only a limited number
of preference matching scores~$s^{b}_{\textbf{u},\textbf{i}'}$ are
computed at the second stage.
Thus, the computational cost of the proposed two-stage ranking method is only
slightly higher than for recommendation based solely on viewing behavior; this is also
discussed in later experiments.

\begin{algorithm}
	\small
	\caption{Two-stage Ranking}
	\label{alg:ts_ranking}
	\KwIn{ $\mathcal{A}_{\rm train}$, $I_{\rm train}$, $I_{\rm test}$, $k$, $\mathbf{u}$}
	\KwOut{${\hat{I}}^{\mathbf{u}}_{\rm test}$ (set consisting of recommended
	programs in $I_{\rm test}$ for user $\mathbf{u}$)}
    $\mathcal{S}^{b} \leftarrow [ ]; \mathcal{S}^{p} \leftarrow []; {\hat{I}}_{\mathbf{u}} \leftarrow []$\\
	 Construct $\mathcal{B^{\mathbf{u}}}$ with Eq.~(\ref{eq:watching_behavior})
	 \\
    \For {each $\mathbf{i}'$ in $I_{\rm test}$}{
		Compute $s^{b}_{\mathbf{u},\mathbf{i'}}$ and $(w, c)$ with
		Eq.~(\ref{eq:watching_behavior_score})\\
        $\mathcal{S}^{b}$.append$\left( \left(\mathbf{i}', (w, c),s^{b}_{\mathbf{u},\mathbf{i}'}\right) \right)$\\
    }
	Sort $\mathcal{S}^{b}$ in ascending order according to
	$s^{b}_{\mathbf{u},\mathbf{i}'}$\\
    \While {$\left(\left|{\hat{I}}^{\mathbf{u}}_{\rm test}\right|< k\right)$}{
    	$(\mathbf{i}', (w, c),s^{b}_{\mathbf{u},\mathbf{i}'}) \leftarrow\mathcal{S}^{b}$.pop() \\
		  Compute $h_{{w}_j,\mathbf{u}}$ (or $h_{\mathbf{u}}$),
		  $h_{\mathbf{i}'}$ and $s^p_{\mathbf{u},\mathbf{i}'}$ with
		  Eqs.~(\ref{eq:text_enc})--(\ref{eq:time_preference_score})\\
        \If{$\mathcal{S}^p \neq \emptyset$  {\bf and} $(w,c) \neq (w_{0}, c_{0})$}{
			 Sort $\mathcal{S}^{p}$ in ascending order according to
			 $s^{p}_{\mathbf{u},\mathbf{i}'}$ \\
             ${\hat{I}}^{\mathbf{u}}_{\rm test}$.append$\left( \mathcal{S}^{p}.{\rm pop()} \right)$ \\
        	$\mathcal{S}^{p} \leftarrow []$ \\	
        }
        $(w_{0}, c_{0}) \leftarrow (w,c)$ \\
        $\mathcal{S}^p$.append$\left( \left(\mathbf{i}', s^{p}_{\mathbf{u},\mathbf{i}'}\right) \right)$
    }
\KwRet{${\hat{I}}^{\mathbf{u}}_{\rm test}$}
\end{algorithm}
\vspace{-0.6cm}

\section{Experiment}
\label{sec:experiment}
\subsection{Dataset and Preprocessing}
We collected user viewing logs, denoted as $D_{\rm raw}$, from a set of set-top
boxes providing linear television service to end users in Japan from Jan 1,
2019 to June 1, 2019.
This period comprises a total of 42,301 unique users and 875,550
distinct programs (denoted as $I_{\rm raw}$), where each user was anonymized using
a hashed ID.
Each log records a channel-switching event for a user, denoted as $d=(u, i, c,
t, \Delta t)$, indicating that user~$u$ switched to channel~$c$ broadcasting 
program~$i$ at UTC timestamp~$t$. Above, $\Delta t$ is the interval between
channel-switching events, which can be considered as the duration of the user's
viewing of the program.
Note that each program was broadcast only once on a channel in the linear TV system.
In addition, each program~$i \in I_{\rm raw}$ was associated with its meta
information (see Definition~\ref{def:meta}). 


Given these data logs~$D_{\rm raw}$ and TV programs~$I_{\rm raw}$, we
first removed viewing logs whose duration was less than
$\Delta_{t_\theta}$ (e.g., 15 minutes in the experiments) to filter
out logs where users were just flipping channels rather than
watching a program.
Formally, we constructed the preprocessed data logs~$D=\{d=(u,i,c,t,\Delta t)|
d\in D_{\rm raw} \wedge \Delta t \geq \Delta t_{\theta}\}$.
We then generated training and testing sets by splitting the processed data 
logs~$D$ based on a timestamp~$t_{\rm split}$ and extracting the logs of 
period~$T_{\rm train}=[t_{\rm split}-\Delta{t_{\rm train}},t_{\rm split})$ for
training (denoted as $D_{\rm train}$) and $T_{\rm test}=[t_{\rm split},t_{\rm
split}+\Delta{t_{\rm test}})$ for testing ($D_{\rm test}$); thus
$I_{\rm train}=\{i\,|\,i\in I_{\rm raw}, \mathbf{s}_i \in T_{\rm train}\}$ and
$I_{\rm test}=\{i\,|\,i\in I_{\rm raw}, \mathbf{s}_i \in T_{\rm test}\}$. 
In our experiments, we constructed four datasets with different values for $t_{\rm split}$
and set $\Delta t_{\rm train}$, $\Delta t_{\rm test}$ to 90 and 7 days,
respectively. 
Table~\ref{table: data stat} contains the dataset statistics.
With user logs in $D_{\rm train}$, the interaction tensor is $\mathcal{A}_{\rm
train}=(a_{\mathbf{u},\mathbf{i},\mathbf{w},\mathbf{c}}) \in
\mathbb{R}^{|U|\times |I_{\rm train}|\times|W| \times |C|}$, where 
    $a_{\mathbf{u},\mathbf{i},\mathbf{w},\mathbf{c}}=
 \sum_{
   (u,i,c,t,\Delta t)\in D_{\rm train}} 
   \mathbbm{1}_{\left\{\left(u,i,w_{\mathcal{T}(t)},c\right)=\left(\mathbf{u},\mathbf{i},\mathbf{w},\mathbf{c}\right)\right\}}.$
Here we consider only user sets $U$ appearing at least once both in $D_{\rm
train}$ and $D_{\rm test}$.
The length of each weekly time slot~$w_i \in W$ was set to 15 minutes by setting
$n$ to $672$.
For validation, we adopted user-implicit feedback extracted from $I_{\rm test}$;
that is, for each user~$\mathbf{u} \in U$, we constructed program 
set~${I}^{\mathbf{u}}_{\rm test}=\{i\,|\,i\in I_{\rm test} \wedge
(\mathbf{u},i,c,t,\Delta t)\in D_{\rm test} \}$ as our ground truth.

\input{data_stat}
\subsection{Baselines and Experimental Setup}
We first built two baselines based on viewing behavior and viewing preferences, the user
characteristics introduced in Sections~\ref{sec:behavior} and~\ref{sec:preference}, respectively. 
Note that for viewing preferences, we tokenized the textual information of each
program using MeCab,\footnote{\url{ https://taku910.github.io/mecab/}} after which we
used the term frequency-inverse document frequency (tf-idf) vectorizer as the
text encoder (see $\mathcal{E}(\cdot)$ in Eq.~(\ref{eq:text_enc})) to represent
items in $I_{\rm train}\bigcup I_{\rm test}$.

In addition, we compared the proposed two-stage ranking approach with a ranking
fusion method that combines the recommendations from the above two
baselines using reciprocal rank fusion (RRF)~\cite{Cormack09}. 
In information retrieval (IR), RRF is a simple but effective method for combining 
document rankings from multiple IR systems.
Formally speaking, given a set of items~$I_{\rm test}$ and a set of ranking
functions~$\mathcal{K}$, where each $\kappa\in \mathcal{K}$ is a function
mapping item~$i \in I_{\rm test}$ to its ranking~$\kappa(i)$, the fusion score
for each item~$i$ is computed as 
$s_{\rm RRF}(i)=\sum_{\kappa\in \mathcal{K}} \frac{1}{\kappa(i)+\eta}$,
where $\eta$ is a hyperparameter to reduce the impact of high-ranking items 
from any of the systems.
With the two ranking functions based on viewing behavior and preferences
(denoted as $\kappa_{b}$ and $\kappa_{p}$, respectively), we have $s_{\rm RRF}(i)=
\frac{1}{\kappa_{b}(i)+\eta}+\frac{1}{\kappa_{p}(i)+\eta}$.
Another baseline is an RRF variant with an additional hyperparameter $\xi$ to control the impact
of two ranking systems, $s_{\rm
RRF}^{\xi}(i)=\frac{\xi}{\kappa_{b}(i)+\eta}+\frac{1-\xi}{\kappa_{p}(i)+\eta}$.

We use the following metrics to evaluate our models: (1) nDCG, (2) precision,
and (3) recall.
For each user~$\mathbf{u} \in U$, we recommend $k=30$ programs among $I_{\rm
test}$ and evaluate model performance with cut-offs $N\in\{10,20,30\}$. 
To fine-tune the hyperparameters for the RRF fusion methods (denoted as RRF
and RRF$^\xi$), we randomly selected 10\% of the users in Dataset~1 as the
development set and searched $\eta$ and $\xi$ in the range of $\{ 1,2,\cdots 100
\}$ and $\{ 0,0.1,\cdots 1 \}$, respectively, for the best performance in
terms of Recall@30.
Additionally, to examine the efficiency of each model,  we
evaluated each model's CPU time cost for inference (seconds/user).\footnote{As the
inference time is measured on a per-user basis, the number of threads does
not impact the measurement.}
For models using viewing preferences (including fusion methods), we computed
and indexed $h_{\mathbf{u},\mathbf{w}}$ (or $h_{\mathbf{u}}$) and
$h_{\mathbf{i'}}$ in advance; thus, for each user at the inference stage, the
computation cost is mainly associated with the dot product between
$h_{\mathbf{u},\mathbf{w}}$ (or $h_{\mathbf{u}}$) and $h_{\mathbf{i'}}$ (for
all programs $\mathbf{i}' \in I_{\rm test}$).
In modeling the viewing behavior, the time cost results are primarily due to the
construction of matrix $\mathcal{B}^{\mathbf{u}}$ and the calculation of
$s^{b}_{\mathbf{u},\mathbf{i}'}$.

\subsection{Quantitative Results}
\begin{table*}[t!]
\centering
\resizebox{1\linewidth}{!}{
\setlength{\tabcolsep}{2.5pt}
\begin{tabular}{clcrrrrrrrrrrr}
  \toprule
 & &&\multicolumn{3}{c}{$N=10$}&\multicolumn{3}{c}{$N=20$}&\multicolumn{3}{c}{$N=30$}& \\
  \cmidrule(lr){4-6} \cmidrule(lr){7-9} \cmidrule(lr){10-12}
 & &Time-aware& nDCG  & Prec. & Recall  & nDCG  & Prec. & Recall  & nDCG  & Prec. & Recall & Time\\ 
  \midrule
\multicolumn{2}{c}{Behavior}&&35.25&33.79&12.26&34.68&30.42&18.39&34.25&27.97&22.91& $\dagger$\textbf{0.27}\\
\multicolumn{2}{c}{Preferences} & \checkmark&13.79&12.91&4.61&13.96&12.13&7.63&14.11&11.45&9.98&1.45\\
 \cdashline{1-13}\\[-0.3cm]
\multirow{6}{*}{Fusion}&\multirow{2}{*}{RRF}&  & 43.82&38.27&12.89&38.87&30.30&17.97&36.41&26.01&21.59&1.45\\
 & &{\checkmark} &
45.99&40.64&13.15&41.02&32.58&18.72&38.25&27.82&22.43&1.46\\
 \cdashline{2-13}\\[-0.3cm]
 &\multirow{2}{*}{RRF$^{\xi}$}& &45.44&39.93&13.69&41.78&33.53&19.66&39.80&29.60&23.83&1.45\\
  & &{\checkmark}&47.79&41.90&13.93&43.35&34.65&19.86&41.13&30.53&24.11&1.46\\
\cdashline{2-13}\\[-0.3cm]
  &\multirow{2}{*}{Two-stage}&&46.32&40.92&13.61&42.61&34.45&19.23&40.54&30.44&23.27&0.30\\
 &&{\checkmark}&$\dagger$\textbf{48.92}&$\dagger$\textbf{43.28}&\textbf{14.12}&$\dagger$\textbf{44.90}&$\dagger$\textbf{36.41}&\textbf{19.98}&$\dagger$\textbf{42.64}&$\dagger$\textbf{32.13}&\textbf{24.20}&0.31\\
   \bottomrule
\end{tabular}%
}
\caption{Recommendation performance.
$\checkmark$ denotes methods using time-aware user preferences, and
$\dagger$ denotes statistical significance at $p < 0.05$.  
}
\label{table:main results}
\vspace{-1cm}
\end{table*}
Table~\ref{table:main results} compares model performance in terms of the
aforementioned metrics and inference time. 
In the table, the best result for each column is in boldface; $\dagger$
denotes statistical significance at $p < 0.05$ (paired $t$-test over four
datasets) with respect to all other methods, and $\checkmark$ indicates 
methods using time-aware user preferences (i.e., $h_{\textbf{u},\textbf{w}}$
in Eq.~(\ref{eq:tpreference})) as opposed to global preferences (i.e.,
$h_{\textbf{u}}$ in Eq.~(\ref{eq:gpreference})).

First, the comparison between the methods using only behavior or preferences
(denoted as Behavior and Preferences, respectively, in the table and hereafter)
is strong evidence that in the TV Rec scenario, user viewing behavior
dominates recommendation performance, which
  underscores the importance   
of putting the
model for viewing behavior at the first stage of the proposed two-stage ranking
approach.
In addition, note that the inference time cost of Behavior is
five times less than that of Preferences. 
On the other hand, as demonstrated in the table, fusing the two user
characteristics significantly boosts ranking performance. 
Specifically, RRF outperforms Behavior in terms of nDCG and Precision by over 7\%
in the low cut-off regions (i.e., $N=\{10, 20\}$).
Tuning the impact of Behavior and Preferences (the second row of RRF$^\xi$ with
$\xi=0.6$) further improves overall ranking performance in terms of nDCG
and precision by over 10\% and recall by over 5\%.
However, both RRF and RRF$^\xi$ include exhaustive dot product computation over
all programs in $I_{\rm test}$, resulting in a time cost per user approximately
equal to that of Preferences.

Table~\ref{table:main results} shows that the proposed two-stage ranking
consistently outperforms other (fusion) methods in terms of efficiency and the 
three evaluation metrics.
Specifically, the method significantly surpasses the strongest baseline
RRF$^{\xi}$ by over 2\% in terms of nDCG and precision 
when modeling user preferences both globally and in a time-dependent fashion; also 
note that time-aware preferences better capture user viewing
preferences and thus yield better performance.
Most importantly, from an efficiency perspective, the time cost of the
two-stage ranking shown in the table is much lower than that of the two fusion
methods and is approximate to that of Behavior, because 
in our method, only a limited number of preference matching scores
involving the dot product operation are computed at the second
stage.
Combining such efficiency and the fact that our method is parameter-free, we
conclude that the proposed method is much more practical than RRF-based
methods. 
\section{Conclusion}
We propose a two-stage ranking approach to fuse two user 
characteristics---viewing behavior and viewing preferences---in a unified manner for TV Rec.
The empirical results on a real-world TV dataset show that our proposed
approach consistently outperforms other baseline methods; more importantly, our
two-stage ranking approach is parameter-free and efficient at inference, making
it applicable and practical to real-world TV Rec scenarios. 

\bibliographystyle{ACM-Reference-Format}
\bibliography{paper}

\appendix

\end{document}

%% file: data_stat.tex
\begin{table*}[t!]
\centering
\resizebox{.7\linewidth}{!}{
\begin{tabular}{ccccccccc}
  \toprule
 Dataset & $t_{\rm split}$ & $|D_{\rm train}|$  & $|I_{\rm train}|$  & $|C|$ & $|U|$& $|I_{\rm test}|$ & $\overline{|{{I}}^{u}_{\rm test}|}$\\
  \midrule
  1 & APR. 01, 2019 &37,859,993  & 257,370 & 173 & 34,392& 31,556 & 53.45\\
  2 & APR. 08, 2019&38,212,364  & 259,514   & 174 & 34,129& 32,504 & 55.47\\
  3 & APR. 15, 2019&38,335,769  & 260,466  & 174 & 33,803& 32,773  & 55.31\\
  4 & APR. 22, 2019&38,415,448  & 261,212 & 177 & 33,817& 33,811 & 55.24\\
  \bottomrule
\end{tabular}%
}
\caption{Data statistics}
\label{table: data stat}
\vspace{-1cm}
\end{table*}